\documentclass[aps,prd,twocolumn,groupedaddress,amssymb,eqsecnum,showpacs]{revtex4}
\usepackage{graphicx}
\begin{document}

\newcommand{\newc}{\newcommand}

\newc{\be}{\begin{equation}}
\newc{\ee}{\end{equation}}
\newc{\ba}{\begin{eqnarray}}
\newc{\ea}{\end{eqnarray}}
\newc{\bea}{\begin{eqnarray*}}
\newc{\eea}{\end{eqnarray*}}
\newc{\D}{\partial}
\newc{\ie}{{\it i.e.} }
\newc{\eg}{{\it e.g.} }
\newc{\etc}{{\it etc.} }
\newc{\etal}{{\it et al.}}
\newcommand{\nn}{\nonumber}

\newc{\ra}{\rightarrow}
\newc{\lra}{\leftrightarrow}
\newc{\no}{Nielsen-Olesen }
\newc{\lsim}{\buildrel{<}\over{\sim}}
\newc{\gsim}{\buildrel{>}\over{\sim}}

\title{Cosmological Effects of Radion Oscillations}
\author{L. Perivolaropoulos}
\email{leandros@physics.upatras.gr} \affiliation{Department of
Physics, University of Patras, \\Department of Physics, University
of Ioannina, Greece}
\author{C. Sourdis}
\email{sourdis@pythagoras.physics.upatras.gr}
\affiliation{Department of Physics, University of Patras, Greece}

\date{\today}

\begin{abstract}

We show that the redshift of pressureless matter density due to
the expansion of the universe generically induces small
oscillations in the stabilized radius of extra dimensions (the
radion field). The frequency of these oscillations is proportional
to the mass of the radion and can have interesting cosmological
consequences. For very low radion masses $m_b$ ($m_b\sim10-100\;
H_0\simeq10^{-32}~eV$) these low frequency oscillations lead to
oscillations in the expansion rate of the universe. The occurrence
of acceleration periods could naturally lead to a resolution of
the coincidence problem, without need of dark energy. Even though
this scenario for low radion mass is consistent with several
observational tests it has difficulty to meet fifth force
constraints. If viewed as an effective Brans-Dicke theory it
predicts $\omega=-1+\frac{1}{D}$ ($D$ is the number of extra
dimensions), while experiments on scales larger than $1mm$ imply
$\omega>2500$. By deriving the generalized Newtonian potential
corresponding to a massive toroidally compact radion we
demonstrate that Newtonian gravity is modified only on scales
smaller than $m_b^{-1}$. Thus, these constraints do not apply for
$m_b>10^{-3}~eV$ (high frequency oscillations) corresponding to
scales less than the current experiments ($0.3mm$). Even though
these high frequency oscillations can not resolve the coincidence
problem they provide a natural mechanism for dark matter
generation. This type of dark matter has many similarities with
the axion.

\end{abstract}

\maketitle

\section{Introduction}
Cosmological theories with submillimeter dimensions have recently
been the focus of several studies
\cite{Arkani-Hamed:1999gq,pap.Cosm,pap.Dim,Csaki:1999mp,pap.Stat}
(see also Ref.\cite{Papantonopoulos:2002ew} for a recent review).
These theories were originally proposed to provide a novel
solution to the hierarchy problem by postulating that the
fundamental Planck mass $M_*$ is close to the $TeV$ scale
\cite{Antoniadis:1990ew,pap.Dim,Randall:1999vf}. This is possible
in theories with extra dimensions \cite{pap.Dim} because Gauss's
law relates the Planck scales of the $4+D$-dimensional theory
$M_*$ and the long distance $4$-dimensional theory $M_{pl}$ by \be
M_{pl}^2 = b_0^D M_*^{D+2} \ee where $b_0$ is the present
stabilized size of the extra dimensions. The size of the extra
dimensions $b(t)$ (the radion field) is usually assumed to be
stabilized to its present value $b_0$ by a `radion stabilizing
potential' $V(b)$. The early evolution of the radion has been well
studied
\cite{Arkani-Hamed:1999gq,Cline:1999ky,Csaki:1999mp,Flanagan:1999dc}
and has been shown to generate an inflationary era with
observationally consistent phenomenology (see also
Ref.\cite{Levin:yw} for earlier studies).

Studies of late evolution of the radion around its potential
minimum $b_0$ have mainly focused on the resolution of the moduli
problem
\cite{Csaki:1999jh,Flanagan:1999dc,Arkani-Hamed:1999gq,Gunther:2000jj}:
{\it How do we dilute the energy in radion oscillations which at
the end of inflation is high enough to overclose the universe?}
Here we assume that the radion is stabilized at $b_0$ either by a
short period of late inflation or by some other mechanism. The
oscillations we consider are mainly induced by redshifting matter
at late times. They can lead to interesting late time cosmological
effects through the coupling of $b(t)$ with the scale factor of
the universe $a(t)$. These effects will be the focus of this
paper. In what follows we will show that
\begin{itemize}
\item
The redshift of matter ($\rho(t) \sim a^{-3}$) generically
produces oscillations of the radion $b(t)$ around its (local)
minimum $b_0$. This distinguishes the radion from an ordinary
minimally coupled to gravity scalar field.
\item
For low radion masses ${\cal{O}}(10-100\; H_0)$ these oscillations
lead to accelerating and decelerating periods of the expansion
factor $a(t)$ and thus to a possible resolution of the coincidence
problem \cite{Turner:2001yu}. However, oscillations in this
frequency regime may be strongly constrained by classical tests of
general relativity and by Casimir force measurements
\cite{Will:2001mx,Long-Price:1999}.
\item
For higher radion masses ($m>10^{-3}~eV$) radion oscillations can
not resolve the coincidence problem but they are consistent with
classical gravity tests and their energy density redshifts like
$a^{-3}$. They can therefore play the role of dark matter.
\end{itemize}

A class of `oscillating physics' models with features similar to
those discussed here has been developed in the context of
non-minimally coupled scalar fields (Brans-Dicke theories) in an
effort to explain the apparent periodicity in pencil beam galaxy
redshift surveys \cite{Broadhurst:be} (low frequency oscillations
of the Brans-Dicke scalar
\cite{Crittenden:1991pb,Morikawa1990,Gonzalez:2001mx}) or to
resolve the discrepancy between measurements of $\Omega_m \simeq
0.1-0.3$ and the inflationary prediction of $\Omega_{tot}=1$ (high
frequency oscillations of the Brans-Dicke scalar
\cite{Accetta:yb,Steinhardt:1994vs,McDonald:1991xm}). The case of
radion oscillations discussed here corresponds to a very specific
type of Brans-Dicke theory where the parameters of the theory are
uniquely determined by the number of extra dimensions $D$.

The structure of this paper is the following: In section II we
derive the cosmological equations assuming $D$ toroidally compact
extra dimensions stabilized by a potential and obtain an
approximate analytic solution for small oscillations around the
potential minimum. We also establish the connection with
Brans-Dicke theory and derive the effective values of the
Brans-Dicke parameters as a function of $D$. In section III we
focus on the case of low radion mass and discuss the basic
cosmological features of the model. The corresponding features for
large radion masses are discussed in section IV. Finally, in
section V we conclude and summarize the main remaining open
issues.

\section{Radion Cosmology}
We consider a flat $4+D$ dimensional spacetime $R^1\times
R^3\times T^D$ with toroidally compact $D$ extra dimensions. The
metric may be written as \be \label{eq101}
g_{MN}=diag[1,-a^2(t)\tilde{g}_{ij},-b^2(t)\tilde{g}_{mn}] \ee
where $M,N$ run from 0 to $D+3$; $ i , j $ run from $1$ to 3 and
$m,n$ run from 4 to $D+3$. Also $a(t)$ is the scale of the
non-compact 3-dimensional flat space (the scale factor of the
universe) and $b(t)$ is the radius of the compactified toroidal
space (the radion field).

The nonzero components of the energy-momentum tensor are given by
\ba \label{eq102} &T_{00}&  =  \rho_{tot} g_{00}\nonumber \\
&T_{ij}& = -p_{a} g_{ij}\\ &T_{mn}&  = -p_{b} g_{mn} \nonumber \ea

The energy density $\rho_{tot}$ and the pressures $p_{a},p_{b}$
are derivable from the internal energy ${\mathcal U} = {\mathcal
U}(a,b)$ as
\ba
    \rho_{tot} = \frac{\mathcal{U}}{\mathcal{V}} ,~~
    p_a = - \frac{a \partial {\mathcal U} /\partial a}{3 \mathcal{V}} ,~~
    p_b = - \frac{b \partial {\mathcal U} /\partial b}{D \mathcal{V}}
\ea where $ \mathcal{V}= a^3(t) \Omega_{D} b^D(t)$ is the volume
of the $(D+3)$-space
($\Omega_D=2\pi^{\frac{D+1}{2}}/\Gamma(\frac{D+1}{2}$)).

Consider now an internal energy of the form
\be
\label{eq100}
{\cal U}=a^3\left(V(b)+\rho\right)
\ee
 $V(b)$ in equation
(\ref{eq100}) is the radion potential which can produce
\cite{Arkani-Hamed:1999gq} sufficient primordial inflation to
solve the horizon, flatness, homogeneity, and monopole problems
and can stabilize $b$ at
\be b_{0}=\left({M_{Pl}^2 \over
M_{*}^{D+2}}\right)^{1/D}
\ee
with a vanishing cosmological
constant. The energy density $\rho=\rho(a)$ in (\ref{eq100}) is
due to mater and radiation.

The action for Einstein gravity in $N=D+4$ dimensions is
\be
S =
 \int d^{N}x \sqrt{-g{^{(N)}}} \biggl( \frac{1}{16\pi
\overline{G}}{\cal {R}}[g] + {\cal L_{\rm matter}} \biggr)
\label{action in N} \ee

The generalized Einstein equations are
\be
\label{eq103} R_{MN} =8\pi \overline{G}\left(T_{MN}-
{{T^K}_K\over{D+2}}g_{MN}\right) \ee
 where
$\overline{G}$ is the $(4+D)$ dimensional gravitational constant.
The gravitational coupling $16\pi G = {1\over b_{0}^D
M_{*}^{D+2}}$ is weak because $b_{0}$ is much greater than the
$(4+D)$-dimensional Planck length $1\over M_{*}$. Thus, the
hierarchy problem is `transferred' to the large extra dimensions.

Using the metric ansatz (\ref{eq101}) and the energy-momentum
components (\ref{eq102}), it is straightforward to obtain the
following set of evolution equations for the scale factors $a$ and
$b$ from equations (\ref{eq103}).
\begin{widetext}
\ba \label{eq104}
&&6 \frac{\dot a^2}{a^2} + D(D-1)\frac{\dot
b^2}{b^2} + 6D \frac{\dot a}{a} \frac{\dot b}{b} = \frac{V +
\rho}{M_*^{D+2} b^D} \nonumber \\
&&\frac{\ddot b}{b} + (D-1) \frac{\dot b^2}{b^2} + 3 \frac{\dot
a}{a}\frac{\dot b}{b} = \frac{1}{M_*^{D+2} b^D} \left(\frac{2
V}{D+2} - \frac{b}{D(D+2)}
\frac{\partial V}{\partial b} + \frac{\rho-3p_{a}}{2(D+2)} \right) \\
&&\frac{\ddot a}{a} + 2\frac{\dot a^2}{a^2} + D \frac{\dot
a}{a}\frac{\dot b}{b} = \frac{1}{M_*^{D+2} b^D} \left(
\frac{b}{2(D+2)} \frac{\partial V}{\partial b} -
\frac{D-2}{2(D+2)} V + \frac{\rho + (D-1) p_{a}}{2(D+2)} \right)
\nonumber \ea
\end{widetext}

In the picture of Ref. \cite{pap.Dim} (ADD) and
\cite{Randall:1999vf} (RS), the matter-radiation energy density is
assumed to be localized on the brane corresponding to the $a(t)$
scale factor. This localized energy density in general distorts
the geometry of the compactified $D$-dimensional space (the bulk),
but as far as the overall properties and the evolution of the
radion are concerned, it is correct to treat the energy density on
the wall as just being averaged over the whole space as done on
the RHS of these equations. This assumption is consistent with the
results of references \cite{Csaki:1999mp,Csaki:1999jh} where it
was shown that the coupling of the radion field to the energy
momentum tensor is given generically through the trace
$(\rho-3p_a)$ plus terms involving the stabilizing potential
$V(b)$.

Equations (\ref{eq104}) have been analyzed in the context of
inflation (away from the stabilization point $b_{0}$) in
\cite{Arkani-Hamed:1999gq} (ADD model).

Similar equations \cite{Csaki:1999mp} arise in the context of
Randall-Sundrum (RS) models \cite{Randall:1999vf}, where the
hierarchy problem is solved using extra dimensions of smaller
sizes at the expense of introducing a non-flat background metric
along the extra coordinates and a pair of branes whose distance
$b(t)$ is stabilized at $b_{0}$ by the potential $V(b)$.

From equations (\ref{eq104}) it is clear that radiation
$(p_{rad}={1\over3} \rho_{rad})$ has no effect on the dynamics of
the radion. This is not the case however for matter. Redshifting
matter plays the role of a driving force and can induce radion
oscillations during both the matter and radiation eras. These
oscillations backreact on the scale factor $a(t)$ through the
effective Friedman equations (\ref{eq104}) and can induce
interesting cosmological effects. In what follows we study these
cosmological effects in the presence of redshifting matter and
radiation neglecting any other possible existence leading to
equations of state with $\rho-3p_a\neq0$.

To study the evolution of the system of the scale factor $a(t)$
coupled to the radion $b(t)$, we focus on the first two equations
of the system (\ref{eq104}) (the last one, given the
energy-momentum conservation and the equation of state,  is not
independent).
 The potential $V(b)$ may be written to lowest order in
 $\delta\equiv{b-b_0\over b_{0}}$ as
\be \label{rad pnt}
 V = \rho_{\rm v}\left({b\over b_0} -1\right)^2 = \rho_{\rm v}\delta^2
 \ee

In order to write this system in dimensionless form we define the
following dimensionless quantities
\ba &&\bar{b} = {b\over b_0} \\
&&\bar{a} = {a\over a_0} \\ &&\overline{m}^2 = {\rho_{\rm v}\over
\rho_{om}} \\ \label{tresc} &&\bar{t} = t \left({\rho_{om} \over
{6M_*^{D+2}b_0^D}}\right)^{1/2} \ea where $a_0$ is the scale
factor at the present time $t=t_0$. Expanding equations
(\ref{eq104}) around the stabilization point $b_0$
($V^{\prime}(b_0)=0$) the radion mass can be read off from the
corresponding linearized equation of motion (see e.g. Ref.
\cite{Arkani-Hamed:1999gq}) and is given in terms of the second
derivative of the potential (\ref{rad pnt}). Up to a constant
factor of order one, the radion mass is \ba
{m_b}^2&=&\frac{b_0^2V^{\prime\prime}}{M_*^{D+2}b_0^D}\equiv
\frac{b_0^2V^{\prime\prime}}{M_{Pl}^2} \nonumber\\
\label{rad-mass} &=&\frac{\rho_{\rm
v}}{{M_{Pl}}^2}=\frac{{\overline{m}}^2\rho_{om}}{{M_{Pl}}^2}\simeq
(10^{-33}{\overline{m}})^2\;\; eV^2 \ea ($\rho_{om}$ is the matter
density at the present time $t_0$). In order to simplify our
notation in what follows, we will omit the bar from the above
dimensionless quantities (except ${\overline{m}}$). Thus, the
system (\ref{eq104}) may be written in dimensionless form as

\begin{widetext}
\ba \label{eq105} &&\frac{\dot a^2}{a^2}
+\frac{D(D-1)}{6}\frac{\dot b^2}{b^2} + D \frac{\dot a}{a}
\frac{\dot b}{b} = \frac{{\overline{m}}^2 (b-1)^2 +
\frac{1}{a^3}}{b^D} \\ \label{eq106} &&\frac{\ddot b}{b} + (D-1)
\frac{\dot b^2}{b^2} + 3 \frac{\dot a}{a}\frac{\dot b}{b} =
\frac{3}{(D+2) b^D}
\left[4{\overline{m}}^2(b-1)^2-\frac{4{\overline{m}}^2
b(b-1)}{D}+\frac{1}{a^3}\right] \ea
\end{widetext}

This set of equations is similar to the cosmological field
equations of a Brans-Dicke theory for specific parameter values.
To identify these parameter values we compare the dimensionally
reduced form of (\ref{action in N})

\be S =
 \int d^{4}x  \frac{\sqrt{-g{^{(4)}}}}{16\pi G}\biggl[b^D{\cal {R}}^{(4)}+D(D-1)b^{D-2}\dot{b}^2 - V_b(b) \biggr]
 \label{action in 4}
 \ee
with the 4-dimensional Brans-Dicke theory of a dynamical Planck
mass \be S = \int d^{4}x  \frac{\sqrt{-g}}{16\pi
G}\biggl[F(\phi){\cal {R}}-Z(\phi)\partial_\mu \phi\partial^\mu
\phi - V_\phi(\phi) \biggr]
 \label{BD action}
 \ee
The identification that needs to be made is \ba &F(\phi)=b^D \\
&-Z(\phi)\dot{\phi}^2=D(D-1)b^{D-2}\dot{b}^2 \ea In the
parameterization $\phi=b^D$ we obtain \ba &F(\phi)=\phi \\
&Z(\phi)=(-1+\frac{1}{D})/\phi \ea which implies
\be
\label{omega} \omega\equiv\frac{ZF}{{F^{\prime}}^2}=-1+\frac{1}{D}
\ee Other parameterizations ({\it e.g.} $\phi=b$ or $Z(\phi)=-1$)
are easily checked to give identical values for $\omega$.

The existence of a finite value of $\omega$ implies a Yukawa
interaction (fifth force) modification of Newton's law
\cite{Steinhardt:1994vs} \be \label{bdnewt} V(r)=-G \frac{M}{r}
\left(1+ \frac{1}{3+2\omega} e^{-m_{\phi}\; r}\right) \ee where
$m_b= \frac{1}{D(D+2)} m_{\phi}$, as can be verified by comparing
the linearized equations of motion for the radion (see
eq.(\ref{eq104})) with the equations of motion for a massive
Brans-Dicke scalar.

Using equation (\ref{omega}) this becomes \be
 V_{m_b}(r) =
-G \frac{M}{r} \left(1+ \frac{D}{D+2} e^{-m_b\; r}\right)
\label{bdnewt1}
 \ee
The Yukawa interaction of equation (\ref{bdnewt1}) is due purely
to radion dynamics while geometrical effects have been integrated
out by the dimensional reduction of equation (\ref{action in 4})
($b_0\to 0$). The Yukawa interaction induced purely by geometrical
effects (fixed $b_0$, $m_b \to \infty$) has been studied in
Ref.\cite{Floratos:1999bv}. The resulting modified Newtonian
potential in that case is \be \label{geomnewt} V_{b_0}(r)=-G
\frac{M}{r} \left(1+ 2D e^{-r/b_0}\right) \ee Combining equations
(\ref{bdnewt1}) and (\ref{geomnewt}) we obtain the total modified
Newtonian potential for a massive toroidally compact radion \be
\label{totnewt} V_{tot}(r)=-G \frac{M}{r}
\left(1+\frac{D}{D+2}e^{-m_b r} + 2D e^{-r/b_0}\right) \ee
Corresponding modified Newtonian potentials can be obtained for
alternative compactification schemes (spherical and Calabi-Yau)
thus generalizing the corresponding results of
Ref.\cite{Floratos:1999bv}.

Fifth force tests on scales larger than $1mm$ (solar system and
terrestrial) imply a constraint \cite{Will:2001mx} $\omega>2500$
for the massless ($V(\phi)=0$) Brans-Dicke model. On scales
$r<<m_{b}^{-1}$ a massive Brans-Dicke model behaves like the
original massless theory. On scales larger than the range of
spatial fluctuations ($r>>m_{b}^{-1}$) the Brans-Dicke scalar
dynamics are frozen by the potential $V(\phi)$ and the model
behaves like Einstein gravity. Therefore fifth force experiments
on these scales do not constrain the massive Brans-Dicke theories
and the constraint $\omega>2500$ is not applicable provided
$m_{b}^{-1}<1mm$ (minimum scale of experimental constrains). A
separate requirement for consistency with fifth force experiments,
coming from the geometric effects of toroidal extra dimensions in
equation (\ref{totnewt}) is $b_0 < 1mm$.

Since we are interested in small radion oscillations, it is
convenient to linearize the system (\ref{eq105}), (\ref{eq106})
setting $\delta=b-1$. Keeping only the dominant terms we obtain
the system \ba &&\frac{\dot{a}^2}{a^2} =
\frac{1}{a^3}-D\frac{\dot{a}}{a}\dot{\delta} \\ \label{inhde}
&&\ddot{\delta}+3\frac{\dot{a}}{a}\dot{\delta} = -m^2 \;
\delta+\frac{3}{D+2}\frac{1}{a^3} \ea where \be
m^2\equiv\frac{12\overline{m}^2}{(D+2)D}\ee
The solution of this
system is well approximated by \ba \delta &\simeq &
\delta_0\frac{\cos(mt+\theta)}{t}\simeq \frac{3}{2}
\delta_0\frac{\cos(mt+\theta)}{a^{3/2}}\\ \label{eq for H}
\frac{\dot{a}}{a} &\equiv & H = -\frac{D}{2}\dot{\delta}+
\left[\left(\frac{D}{2}\dot{\delta}\right)^2+\frac{1}{a^3}\right]^{1/2}
\ea where $\delta_0$ is determined by the initial conditions at
the time $t_i$ when the oscillations start. For radion
oscillations induced purely by the redshifting matter we keep only
the particular solution of the inhomogeneous differential equation
(\ref{inhde}) and find
\be
\label{matsol} \delta_0 \simeq
-\frac{4\pi}{3(D+2)m},\;\;\;\;\theta=0 \ee

In the following section we briefly discuss the cosmological
effects of this solution for very low radion mass.

\section{Low Radion Mass}
For $m\sim{\cal{O}}(10-100\; H_0)\simeq10^{-32}~eV$ the fifth
force terrestrial and solar system constrains apply
($\omega>2500$) and due to equation (\ref{omega}) the model is not
consistent with them at the classical level. A possible way out of
this constraint are modifications of the effective Brans-Dicke
parameters due to quantum effects \cite{Albrecht:2001xt}.

Nevertheless, the model has several interesting features which are
worth discussing before proceeding to the phenomenologically more
relevant case of large $m$.

For radion masses comparable to $H_0$, the dominant corrections
come from the linear term in $\dot{\delta}$ which does not average
out to zero. Thus we obtain \be \label{perh} H =
\overline{H}-\frac{D}{2}\dot{\delta}\simeq
\overline{H}\biggl(1+\xi\; m\; \sin(mt+\theta)\biggr) \ee where
$\overline{H}=\frac{1}{a^{3/2}}$ is the unperturbed Hubble
parameter and $\xi,\theta$ are constants depending on the initial
conditions of the radion oscillations and the number of extra
dimensions $D$ ($\xi\equiv 3\delta_0 D/4$).

For \be \label{accond} m\; \xi\geq\frac{1}{2}\ee equation
(\ref{perh}) implies periodically accelerating Hubble expansion.
At maximum acceleration the expansion factor is \be a(t)\sim
t^{2(1+m\xi )/3} \ee

For radion oscillations induced purely by redshifting matter (the
source term $\frac{1}{a^3}$) we may use (\ref{matsol}) to obtain
\be \label{mxmat} m\;\xi\simeq\frac{\pi D}{(D+2)}>\frac{1}{2}\ee
and therefore these oscillations are sufficient to produce periods
of accelerating expansion.

The periods of acceleration of the scale factor consist a very
interesting feature for the following reasons:
\begin{enumerate}
\item
\textit{No Dark Energy:} They can potentially explain the
acceleration of the scale factor observed in the recent SNe Ia
data \cite{Perlmutter:1998np} without the requirement of any form
of dark energy. Alternative attempts to achieve accelerating
expansion without the need of dark energy can be found in
\cite{Freese:2002sq,Sahni:2002dx}.
\item
\textit{Coincidence problem:} They can resolve the coincidence
problem by inducing other accelerating and decelerating periods in
the past. Alternative resolutions of the coincidence problem based
on oscillating energy of minimally coupled scalars with
oscillating potentials \cite{Dodelson:2001fq,Griest:2002cu} or
quintessence induced by extra dimensions
\cite{Pietroni:2002ey,Albrecht:2001xt} have also been recently
proposed.
\item
\textit{Consistency:} They are consistent with nucleosynthesis
constrains assuming that they start at $t_i\simeq m^{-1} > t_{eq}$
\textit{i.e.} much later than the time of nucleosynthesis (this is
the behavior expected for oscillating scalars in an expanding
universe \cite{Turner:1983he}).
 The model is also consistent with
observational constraints other than the fifth force
\cite{inprog}: The age of the universe is longer compared to the
CDM model without cosmological constant (SCDM). Structure
formation is mildly affected. The first Doppler peak of the cosmic
background radiation is shifted only slightly and remains
consistent with experimental results. The time dependence of
Newton's constant constrained by measurements of spin-down rate of
pulsars, imposes some restrictions on the parameter $\theta$ which
are more difficult to meet for a {\it large number} of extra
dimensions.
\item
\textit{Predictiveness:} They make the prediction of deceleration
at high redshifts at a rate higher than that of non-oscillating
models.
\item
\textit{Naturalness:} They are well motivated without the
requirement of extra scalar fields or potentials.
\item
\textit{Periodicity of galaxy distribution:} The north-south
pencil beam survey of Ref.\cite{Broadhurst:be} suggests an
apparent periodicity in the galaxy distribution. The number of
galaxies as a function of redshift seems to clump at regularly
spaced intervals of $128h^{-1} Mpc$. Recent simulations
\cite{Yoshida:2001} have indicated that this regularity has a
priori probability  less than $10^{-3}$ in CDM universes with or
without a cosmological constant. This suggests a new cosmological
puzzle. Low mass radion oscillations induce a modulation in the
galaxy redshift count by the factor \be
\frac{dz}{dz_0}=\frac{H}{\overline{H}}=1+\xi\; m\; \sin(mt+\theta)
\ee Such oscillations of $\frac{dz}{dz_0}$ can explain the peaks
in the survey of Ref.\cite{Broadhurst:be} provided that their
amplitude is larger than $1/2$ \cite{Hill:1990uy}. This condition
is identical with the condition (\ref{accond}) required to have
periods of accelerated expansion for solving the coincidence
problem. It may be shown \cite{inprog} that the value of $m$
required to induce a periodicity of $128 h^{-1} Mpc$ for $D=2$ is
$m\simeq 100$ which corresponds to $m_b \simeq 10^{-31}eV$.
\end{enumerate}

Despite these interesting features, the main drawback for this
range of radion mass is the possible inconsistency of the model
with fifth force constraints, as discussed above.

\section{High Radion Mass}
For $m_b>10^{-3}eV>>\overline{H}_0$ ($m\agt 10^{30}$), the
constraint $\omega>2500$ applies only for experiments on scales
$r\alt 1mm$. Such experimental constraints are not available at
present and therefore radion oscillations are experimentally
allowed for this mass range. In this case, the linear term in
equation (\ref{eq for H}) can be ignored since it averages out to
zero. Thus we obtain \ba
H^2&\simeq&\biggl[\frac{1}{a^3}+\left(\frac{D}{2}\dot{\delta}\right)^2\biggr]\nonumber\\
 &\simeq&\left(\frac{1}{a^3}+\frac{4\;\xi^2m^2\sin^2(mt+\theta)}{9\;t^2}\right)\\
\label{newdm}
&=&\frac{1}{a^3}\biggl(1+\frac{\Omega_{0r}}{\Omega_{0m}}\biggr)
\ea where in the last equation we have considered averaging over
time, used (\ref{tresc}) and \ba t_0^{-2} &=&
\frac{9}{4}\overline{H}_0^2 = \frac{9~\rho_c(t_0)}{4~\rho_m(t_0)}
=
\frac{9}{4\Omega_{0m}}\\ \label{om0r} \Omega_{0r} &\equiv &
\frac{1}{2}\xi^2m^2 \ea

Therefore equation (\ref{newdm}) implies that in this mass range
the oscillating radion can play the role of a new dark matter
component redshifting as $a^{-3}$ with present relative energy
density \be \Omega_{0r} =
\biggl(\frac{D}{2}\dot{\delta}(t_0)\biggr)^2 \ee

For radion potentials of the form $V(\delta)\sim \delta^{2n}$ with
$n>1$ the radion oscillation energy redshifts slower than $a^{-3}$
and can result to accelerating expansion of the universe. This
effect was studied in the context of a minimally coupled scalar
field in Ref.\cite{Sahni:1999qe} where the resulting dark energy
was termed ``Frustrated Cold Dark Matter".

For radion oscillations induced purely by redshifting matter (see
eqs. (\ref{mxmat}) and (\ref{om0r})) we obtain $\Omega_{0r}$ of
${\cal{O}}(1)$ independent of $m$. Setting $\Omega_{0r} =
{\cal{O}}(1)$ we find an amplitude of the radion oscillation \be
\delta = \frac{\dot{\delta}}{m}={\cal{O}}(m^{-1}) \ee For $m_b
\agt 10^{-3} eV$ we have \be \frac{\delta G}{G}\simeq \delta \alt
{\cal{O}}\left(10^{-30}\right) \ee which is consistent with
nucleosynthesis and is not likely to produce observable
astrophysical or cosmological effects.

The high radion mass ($m_b > 10^{-3} eV$) has therefore three
important effects
\begin{itemize}
\item
It allows for a significant contribution of the oscillating energy
to the energy density of the universe.
\item
It suppresses the amplitude of the oscillations making them
compatible with terrestrial measurements of $G$, nucleosynthesis
constraints and stellar evolution.
\item
It confines the fifth force type effects to scales less than
${\cal{O}}(m_b^{-1})\alt 1mm$ making them compatible with tests
for intermediate range forces and solar system tests of general
relativity \cite{Will:2001mx,Long-Price:1999}.
\end{itemize}

\section{Conclusions - Open Issues}
Our conclusions can be summarized as follows
\begin{itemize}
\item
Radion oscillations are generically induced at late times by
redshifting matter.
\item
For low radion masses ($m\simeq10-100\; H_0$) these oscillations
could provide a solution to two important cosmological problems:
\textit{the coincidence problem} (why do we live at the special
time when the universe's expansion begins to accelerate) and
\textit{the apparent periodicity of galaxy distribution} with
spatial period $\simeq128h^{-1}Mpc$. However, fifth force
constraints based on solar system and terrestrial observations may
not be consistent with this range of radion masses.
\item
For high radion masses ($m_b>10^{-3}eV$) radion oscillations are
consistent with fifth force and other constraints and they can
provide the source of a new type of dark matter which has many
similarities with axions (they are both a result of oscillating
scalars).
\end{itemize}

Open issues that require further study are the following :
\begin{itemize}
\item
Can quantum effects modify the effective Brans-Dicke parameters of
low mass radion oscillations, making them consistent with fifth
force constraints while still allowing the resolution of the
coincidence and galactic periodicity problem?
\item
What are the clustering properties of the oscillating radion dark
matter?
\item
What experimental or observational tests could detect the low
amplitude - high frequency of the radion and the corresponding
Newton's constant $G$?
\end{itemize}
{\bf Acknowledgements:} We thank I. Bakas for several useful
discussions and for critically reading the manuscript. We
acknowledge support by the University of Patras through the `C.
Caratheodory' research grants No. 2793 and No. 2453. This work was
supported in part by a network of the European Science Foundation
and by the European Research and Training networks ``Superstring
Theory" (HPRN-CT-2000-00122) and ``The Quantum Structure of
Space-time" (HPRN-CT-2000-00131).


\begin{thebibliography}{99}

\bibitem{Arkani-Hamed:1999gq}
N.~Arkani-Hamed, S.~Dimopoulos, N.~Kaloper and J.~March-Russell,
Nucl.\ Phys.\ B {\bf 567}, 189 (2000) [arXiv:hep-ph/9903224].

\bibitem{pap.Cosm}
N. Kaloper and A. Linde,
{\it Phys. Rev.}  {\bf{D 59}} {\it  101303 (1999)}
[hep-th/9811141]; N.~Arkani-Hamed, S.~Dimopoulos, G.~R.~Dvali and
N.~Kaloper,
Phys.\ Rev.\ Lett.\  {\bf 84} 586 (2000) [arXiv:hep-th/9907209];
R. N. Mohapatra, A. Perez-Lorenzana, C. A. de S. Pires, {\it
Int.J.Mod.Phys.} {\bf A16} {\it 1431 (2001)}, [hep-ph/0003328]; N.
Kaloper.
{\it Phys. Rev.} {\bf{D60}} {\it 123506 (1999)} [hep-th/9905210];
P. Kanti, I. I. Kogan, K. A. Olive and M. Pospelov,
{\bf{B468}} {\it  31 (1999)} [hep-ph/9909481].

\bibitem{pap.Dim}
N. Arkani-Hamed, S. Dimopoulos and G. Dvali,
{\it Phys. Lett.} {\bf{B 429}} {\it  263 (1998)} [hep-ph/9803315],
{\it Phys. Rev.}{\bf{D 59}} {\it  086004 (1999)} [hep-ph/9807344];
I.~Antoniadis, N.~Arkani-Hamed, S.~Dimopoulos and G.~R.~Dvali,
Phys.\ Lett.\ B {\bf 436}  257 (1998) [arXiv:hep-ph/9804398].

\bibitem{Csaki:1999mp}
C.~Csaki, M.~Graesser, L.~Randall and J.~Terning,
Phys.\ Rev.\ D {\bf 62}, 045015 (2000) [arXiv:hep-ph/9911406];
P.~Kanti, I.~I.~Kogan, K.~A.~Olive and M.~Pospelov,
Phys.\ Rev.\ D {\bf 61} 106004 (2000) [arXiv:hep-ph/9912266].


\bibitem{pap.Stat}
G. Dvali and S. H. H. Tye,
{\it Phys. Lett.} {\bf{B450}} {\it  72 (1999)} [hep-ph/9812483];
H. B. Kim and H. D. Kim,
[hep-th/9909053]; C. Csaki, M. Graesser, C. Kolda and J. Terning,
{\it Phys. Lett.} {\bf{B462}} {\it  34 (1999)} [hep-ph/9906513];
J. Cline, C. Grojean and G. Servant,
{\it
 Phys. Rev. Lett. } {\bf 83} {\it  4245 (1999)}, [hep-ph/9906523];

\bibitem{Papantonopoulos:2002ew}
E.~Papantonopoulos,
arXiv:hep-th/0202044.

\bibitem{Antoniadis:1990ew}
I.~Antoniadis,
Phys.\ Lett.\ B {\bf 246}  377 (1990).

\bibitem{Randall:1999vf}
L.~Randall and R.~Sundrum, Phys.\ Rev.\ Lett.\  {\bf 83}, 4690
(1999) [arXiv:hep-th/9906064]; Phys.\ Rev.\ Lett.\  {\bf 83},
  3370 (1999) [arXiv:hep-ph/9905221].

\bibitem{Cline:1999ky}
J.~M.~Cline,
Phys.\ Rev.\ D {\bf 61}, 023513 (2000) [arXiv:hep-ph/9904495].

\bibitem{Flanagan:1999dc}
E.~E.~Flanagan, S.~H.~Tye and I.~Wasserman,
Phys.\ Rev.\ D {\bf 62}, 024011 (2000) [arXiv:hep-ph/9909373].

\bibitem{Levin:yw}
J.~J.~Levin,
Phys.\ Lett.\ B {\bf 343}, 69 (1995) [arXiv:gr-qc/9411041].

\bibitem{Csaki:1999jh}
C.~Csaki, M.~Graesser, C.~Kolda and J.~Terning,
Phys.\ Lett.\ B {\bf 462}, 34 (1999) [arXiv:hep-ph/9906513];

\bibitem{Gunther:2000jj}
U.~Gunther and A.~Zhuk,
Phys.\ Rev.\ D {\bf 61} 124001  (2000) [arXiv:hep-ph/0002009].

\bibitem{Sahni:1999qe}
V.~Sahni and L.~M.~Wang,
Phys.\ Rev.\ D {\bf 62} 103517 (2000) [arXiv:astro-ph/9910097].

\bibitem{Sahni:2002dx}
V.~Sahni and Y.~Shtanov,
arXiv:astro-ph/0202346.

\bibitem{Turner:2001yu}
M.~S.~Turner,
arXiv:astro-ph/0108103; M.~S.~Turner,
Phys.\ Scripta {\bf T85}, 210 (2000) [arXiv:astro-ph/9901109].

\bibitem{Will:2001mx}
C.~M.~Will,
Living Rev.\ Rel.\  {\bf 4}, 4 (2001) [arXiv:gr-qc/0103036].


\bibitem{Long-Price:1999}
J.~C.~Long, H.~W.~Chan and J.~C.~Price
Nucl.\ Phys.\ B {\bf 539},  23 (1999) [arXiv:hep-ph/9805217].

\bibitem{Broadhurst:be}
T.~J.~Broadhurst, R.~S.~Ellis, D.~C.~Koo and A.~S.~Szalay,
Nature {\bf 343},  726 (1990).

\bibitem{Crittenden:1991pb}
R.~G.~Crittenden and P.~J.~Steinhardt,
Astrophys.\ J.\  {\bf 395},  360 (1992) [arXiv:astro-ph/9812133].

\bibitem{Morikawa1990}
M. Morikawa,
Astrophys.\ J.\ Lett.  {\bf 362}, L.37 (2001)

\bibitem{Gonzalez:2001mx}
J.~A.~Gonzalez, H.~Quevedo, M.~Salgado and D.~Sudarsky,
Phys.\ Rev.\ D {\bf 64}, 047504 (2001) [arXiv:astro-ph/0103314].

\bibitem{Accetta:yb}
F.~S.~Accetta and P.~J.~Steinhardt,
Phys.\ Rev.\ Lett.\  {\bf 67}, 298 (1991).

\bibitem{Steinhardt:1994vs}
P.~J.~Steinhardt and C.~M.~Will,
Phys.\ Rev.\ D {\bf 52}, 628 (1995) [arXiv:astro-ph/9409041].

\bibitem{McDonald:1991xm}
J.~McDonald,
Phys.\ Rev.\ D {\bf 44}, 2325 (1991) [Erratum-ibid.\ D {\bf 45},
1433 (1991)];
Phys.\ Rev.\ D {\bf 48}  2462 (1993).


\bibitem{Albrecht:2001xt}
A.~Albrecht, C.~P.~Burgess, F.~Ravndal and C.~Skordis,
Phys.\ Rev.\ D {\bf 65} 123507 (2002) [arXiv:astro-ph/0107573].

\bibitem{Floratos:1999bv}
E.~G.~Floratos and G.~K.~Leontaris,
Phys.\ Lett.\ B {\bf 465}, 95 (1999) [arXiv:hep-ph/9906238];
A.~Kehagias and K.~Sfetsos,
Phys.\ Lett.\ B {\bf 472}, 39 (2000) [arXiv:hep-ph/9905417].

\bibitem{Perlmutter:1998np}
S.~Perlmutter {\it et al.}  [Supernova Cosmology Project
Collaboration],
Astrophys.\ J.\  {\bf 517}, 565 (1999) [arXiv:astro-ph/9812133].

\bibitem{Freese:2002sq}
K.~Freese and M.~Lewis,
Phys.\ Lett.\ B {\bf 540}  1 (2002) [arXiv:astro-ph/0201229].

\bibitem{Griest:2002cu}
K.~Griest,
arXiv:astro-ph/0202052.

\bibitem{Dodelson:2001fq}
S.~Dodelson, M.~Kaplinghat and E.~Stewart,
Phys.\ Rev.\ Lett.\  {\bf 85}, 5276 (2000)
[arXiv:astro-ph/0002360].

\bibitem{Pietroni:2002ey}
M.~Pietroni,
arXiv:hep-ph/0203085.

\bibitem{Turner:1983he}
M.~S.~Turner,
Phys.\ Rev.\ D {\bf 28}, 1243 (1983).

\bibitem{inprog}
L. Perivolaropoulos and C. Sourdis, unpublished.

\bibitem{Yoshida:2001}
N. Yoshida {\it et al.},
MNRAS  {\bf 325}, 803 (2001) [arXiv:astro-ph/0011212].

\bibitem{Hill:1990uy}
C.~T.~Hill, P.~J.~Steinhardt and M.~S.~Turner,
Astrophys.\ J.\  {\bf 366}, L57 (1991).



\end{thebibliography}
\end{document}